\documentclass[aip,preprint]{revtex4-1}%linenumbers
\usepackage{graphicx}% Include figure files

\draft % marks overfull lines with a black rule on the right

\begin{document}

% Use the \preprint command to place your local institutional report number 
% on the title page in preprint mode.
% Multiple \preprint commands are allowed.
%\preprint{}

%在我们的分析中只用到了高温数据

\title{Numerical simulation of two-component attractive Fermi gases based on parametrized partition function}

\author{Yunuo Xiong}
\email{xiongyunuo@hbpu.edu.cn}
\affiliation{Center for Fundamental Physics and School of Mathematics and Physics, Hubei Polytechnic University, Huangshi 435003, China}

\author{Hongwei Xiong}
\email{xionghongwei@hbpu.edu.cn}

\affiliation{Center for Fundamental Physics and School of Mathematics and Physics, Hubei Polytechnic University, Huangshi 435003, China}

\affiliation{College of Science, Zhejiang University of Technology, Hangzhou 310023, China}

\affiliation{Wilczek Quantum Center, Shanghai Jiao Tong University, Shanghai 200240, China}

\begin{abstract}
The zero-temperature and finite-temperature thermodynamics of two-component Fermi gases with finite-range attractive interaction suffer from fermion sign problem, which seems like an insurmountable problem in exact numerical simulations. In a recent work, we find a reliable method to simulate the thermodynamic properties of single-component Fermi gases for both noninteracting and repulsively interacting cases based on the method of parametrized partition function and the $\xi_E$ curve of constant energy. In the present work, this method is generalized to two-component Fermi gases with finite-range attractive interaction, which shows clearly that our method has good chance to apply to various Fermi systems. From the simulated heat capacity, we find a peak at the temperature below the Fermi temperature which implies the pairing of fermions with different spin. At high temperature, the simulated heat capacity approaches the classical value. The reasonable result in this work validates the application of our method to attractive cases, which implies a wide range of applications, from nuclear physics, BCS-BEC crossover, superconductivity, to neutron star, etc..
\end{abstract}

% insert suggested PACS numbers in braces on next line

\maketitle

\section{Introduction}

The numerical simulation of finite-range attractive Fermi gases from zero temperature to high temperature by exact numerical simulation based on path integral formalism \cite{barker} is still an unsolved problem because of fermion sign problem \cite{ceperley,Alex,troyer,loh,lyubartsev,vozn,Science,Wu,Umrigar,Li,Wei,Yao2,HirshbergFermi,DornheimMod,Xiong2,XiongFSP,XiongPara,Dornheim}. This question is so hard that in 2020, with an ingenious method \cite{HirshbergFermi} of an auxiliary repulsive interaction and extrapolation, only three spin polarized noninteracting fermions at low temperature was successfully simulated by path integral molecular dynamics. Because the attractive interaction between fermions in different spin will make the fermion sign problem become more severe, and it is impossible to extrapolate the energy from repulsive to attractive interaction, it seems that the method in Ref. \cite{HirshbergFermi} can not be applied to this situation corresponding to a wide range of interesting physics. Nevertheless, many methods to overcome fermion sign problem are developed, and applied to a wide range of physics \cite{nodes,Helium,Militzer,Mak,Blunt,Malone,Schoof1,Schoof2,Schoof3,Yilmaz,PB1,PB2,Joonho, WDM}.

In our recent work \cite{Xiong-xi}, with the method of parametrized partition function \cite{XiongFSP,XiongPara} and the $\xi_E$ orbit with constant energy, we succeed in simulating the spin polarized fermions for noninteracting case up to fifty particles and repulsively interacting situation up to twenty particles. In Ref. \cite{Xiong-xi}, with a general analysis, we argue that the idea based on $\xi_E$ orbit can be applied for general Fermi system. Nevertheless, we need to carry out numerical simulation for fermions with attractive interaction to justify our argument.

It is the purpose of the present work to extend our previous method \cite{Xiong-xi} to consider the thermodynamics of two-component fermions with finite-range attractive interaction. We combine our previous work on path integral molecular dynamics for two-component bosons \cite{Xiong4} with the new parametrized path integral formulation in order to extract the thermodynamic properties of two-component fermions. In particular, we study the temperature dependence of energy for a system of two-component fermions with divergent s-wave scattering length through numerical simulation, and find the complete heat capacity curve. We observe a bump in the heat capacity signaling a phase crossover, due to the pairing effect between fermions of different spin, we also compare our findings with the corresponding Bose system where a phase transition occurs at a critical temperature value.

\section{A brief introduction to parametrized partition function and $\xi_E$ orbit}

Because of the Pauli exclusion principle for fermions, for the zero-temperature (ground-state) Fermi system, the corresponding Bose system of the same energy is at medium or high temperature. This observation suggests the following duality property:

\textit{There exists a duality between Bose system at medium or high temperature and the corresponding Fermi system at low temperature.}

This observation suggests a parametrized partition function $Z(\xi,\beta)$ ($\beta=1/k_BT$) to study the thermodynamics of fermions, firstly proposed in Ref. \cite{XiongFSP}. In all our calculations, we use the convention of $k_B=1$. Here, the real parameter $\xi$ interpolates continuously between the fermions with $\xi=-1$ and bosons with $\xi=1$. Once we use the path integral formalism to express this parametrized partition function as a high dimensional integral, we may use the following formula to calculate the energy
\begin{equation}
E(\xi,\beta)=-\frac{\partial Z(\xi,\beta)/\partial \beta}{Z(\xi,\beta)}.
\end{equation} 
However, for $\xi=-1$, the partition function $Z(\xi,\beta)$ is in fact the sum of many terms with half of them positive, half of them negative. This so called fermion sign problem is so severe that at low temperature, roughly speaking, we have the following sign factor
\begin{equation}
s=\frac{Z(\xi=-1,\beta)}{Z(\xi=1,\beta)}\sim e^{-\beta N}.
\label{sign}
\end{equation}
Here $N$ is the total particle number. The energy of fermions is
\begin{equation}
E=\frac{A}{s}.
\end{equation}
To obtain $E$, we need to calculate $A$ and $s$ separately, with path integral Monte Carlo or path integral molecular dynamics.
The exponential decay shown by Eq. (\ref{sign}) makes the direct calculation of the energy for fermions exponentially hard by both path integral Monte Carlo and path integral molecular dynamics. Even for $N=3$, as shown in a recent work \cite{HirshbergFermi}, it is still hard to calculate accurately the energy of fermions for $T<1/6$ , while not possible for zero temperature.

\begin{figure}[htbp]
\begin{center}
 \includegraphics[width=0.9\textwidth]{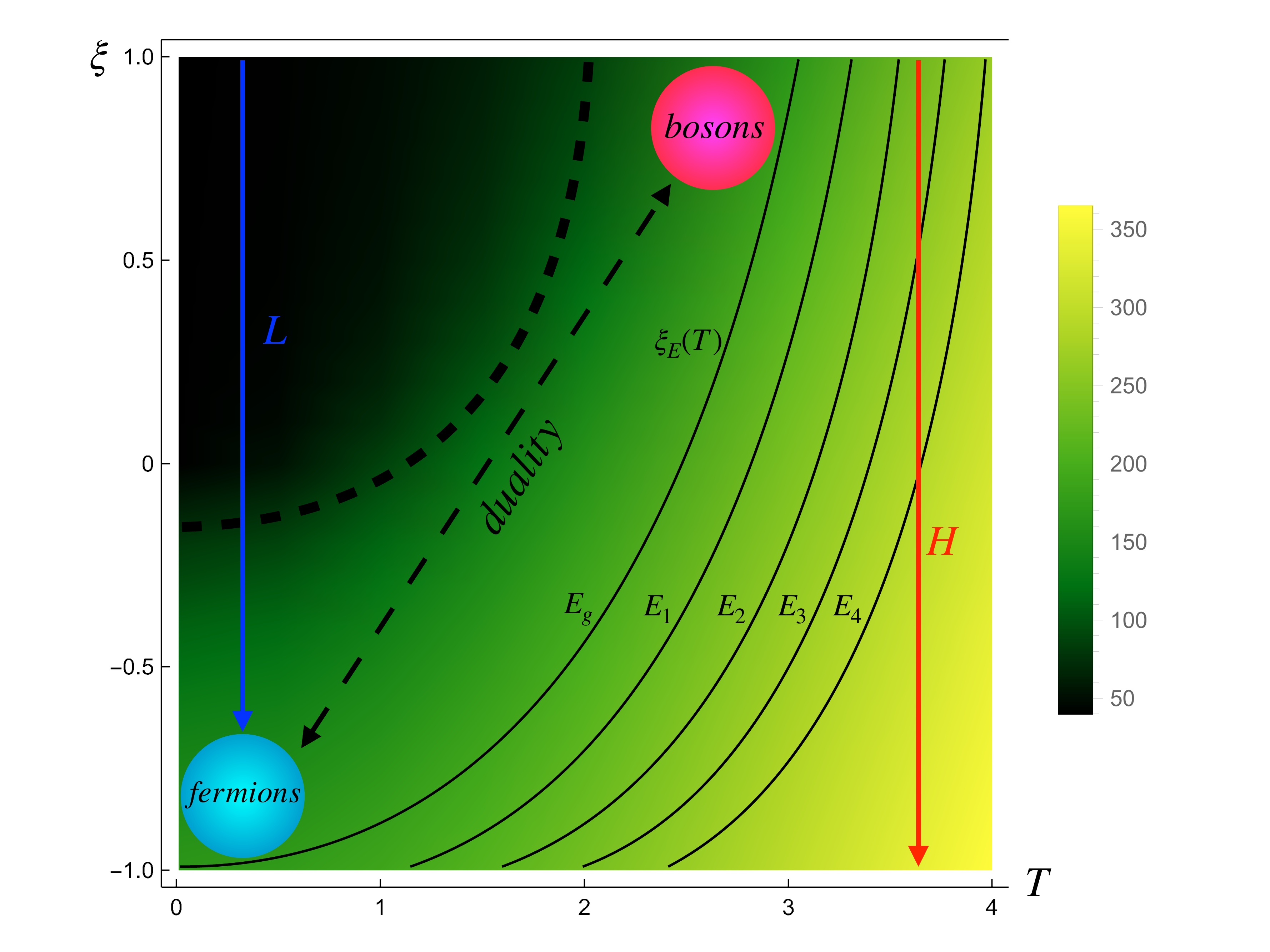} 
\caption{\label{illustration}  The complete contour map for the energy of $50$ noninteracting particles in a three-dimensional harmonic trap. In the black region, the energies are almost constant, and $E(\xi,T)$ is non-analytical there. Extrapolation along the vertical line $H$ works, but along the line $L$ fails. In contrast, the constant energy contour $\xi_E(T)$ bypasses the black region, and we can map the ground state energy of fermions into high temperature bosonic energy through the curve labelled by $E_g$. Finite temperature energies of fermions can also be obtained via constant energy contours, such as those labelled $E_1$, $E_2$, $E_3$, $E_4$.}
\end{center}
\end{figure}

In Fig. \ref{illustration}, for $40$ two-component noninteracting particles in a two-dimensional harmonic trap, we give the energy contour of the function $E(\xi,\beta)$ by numerical calculation with the analytical expression of grand canonical ensemble \cite{XiongPara}. 
\begin{equation}
N=2\sum_{\textbf n}\frac{1}{e^{\beta({\epsilon({\textbf n})-\mu})}-\xi},
\label{chemical}
\end{equation}

\begin{equation}
E(\xi,\beta,N)=2\sum_{\textbf n}\frac{\epsilon(\textbf n)}{e^{\beta({\epsilon({\textbf n})-\mu})}-\xi}.
\label{energyc}
\end{equation}
Here, $\epsilon({\textbf n})$ is the single-particle eigenenergy of the system. The factor $2$ is due to two spin states of fermions. With given parameters of $N,\beta$, and $\xi$,  from Eq. (\ref{chemical}), we can get the chemical potential $\mu(\xi,\beta,N)$. Using further Eq. (\ref{energyc}), we can get the energy $E(\xi,\beta,N)$ in grand canonical ensemble.

Because $E(\xi,\beta)$ is a  monotonic function of both $\xi$ and $\beta$, it is the idea of Ref. \cite{XiongFSP} that we may predict the energy of fermions by firstly calculating $E(\xi,\beta)$ with different positive $\xi$ without fermion sign problem for the same temperature. At medium temperature or high temperature, we may predict reliably the energy of fermions by extrapolation,  shown by the red line "H" in Fig. \ref{illustration}. However, for low temperature, this extrapolation method becomes unreliable. At zero temperature, for noninteracting particles, $E(\xi\geq 0,T=0)$ is a constant, while it increases to the value of the energy of fermions in the region of $\xi< 0$. It is clear that $E(\xi,T=0)$ is not an analytical function of $\xi$ at all for this situation, which proves the invalidity of the extrapolation method along constant temperature shown by the blue line "L" in Fig. \ref{illustration}.

From the structure of the energy contour shown by Fig. \ref{illustration}, the dashed black line encloses a black region where the energy only varies slightly. Hence, any constant-temperature line passing through this black region can not predict correctly the energy of fermions, while outside this region we have good chance with the analysis of the constant-temperature line. 

It is interesting to notice that outside the whole black region, $E(\xi,T)$ has relatively simple monotonic behavior about both $\xi$ and $T$. If we consider the $\xi_E(T)$ curve with constant energy, this curve connects continuously the bosons and fermions without passing through the black region anymore. This verifies the above conjecture that there exists a duality between low-temperature Fermi system and the corresponding high-temperature (or medium temperature) Bose system.

The task now becomes the determination of the $\xi_E(T)$ orbit for the energy larger than the ground state energy $E_g$ of fermions. General analysis shows that the equation of $\xi_E(T)$ orbit can be written as 
\begin{equation}
\xi_E+\sum_{n\geq 2}d_n(E)\xi_E^n=a(E)+b(E)T^2+\sum_{n>2}c_n(E)T^n.
\end{equation}
Here, the coefficients $d_n$ and $c_n$ represent the perturbation series expansion. The absence of the linear term of $T$ on the right hand side of the above equation is due to the property that $\lim_{T\rightarrow 0}\frac{\partial E(\xi,T)}{\partial T}=0$. It is clear that the information of the energy $E(\xi\geq 0,T)$ can give us good chance to determine these coefficients, i.e., the $\xi_E(T)$ orbit. Once these coefficients for a given energy is obtained, by setting $\xi_E=-1$ in the above equation, we get the temperature of fermions with this energy. Usually, we calculate the energy for a given temperature. Now, we predict the temperature of a Fermi system with a given energy in advance.

In our previous work \cite{Xiong-xi}, we use the following perturbation expansion expression
\begin{equation}
\xi_E+a(E)\xi_E^2\approx b(E)+ c(E) T^2+d(E) T^3,
\label{xiexample}
\end{equation}
to study the thermodynamics of two-dimensional noninteracting spin-polarized fermions up to $50$ particles. Good agreement is found, compared with the analytical result of grand-canonical ensemble. For two-dimensional spin-polarized fermions up to $20$ particles having Coulomb repulsive interaction, we also give reasonable simulation of the thermodynamics. The present work will develop the idea introduced in this section to two-component Fermi system with finite-range attractive interaction, so that we may have a wider application of our method.

\section{Parametrized partition function for two-component fictitious particles and path integral molecular dynamics}

For any quantum system, the partition function is
\begin{equation}
Z(\beta)=Tr(e^{-\beta\hat H}).
\end{equation}
The parametrized partition function for single-component identical particles with a parameter $\xi$ can be written as \cite{XiongFSP,XiongPara,Xiong-xi}
\begin{equation}
Z(\xi,\beta)\sim\sum_{p\in S_N}\xi^P\int d\textbf{r}_1d\textbf{r}_2\cdots d\textbf{r}_N\left<p\{\textbf{r}\}|e^{-\Delta\beta\hat H}\cdots e^{-\Delta\beta\hat H}|\{\textbf{r}\}\right>.
\label{Xipartition}
\end{equation}
$S_N$ represents the set of $N!$ permutation operations. The factor $\xi^P$ is due to the exchange effect of identical particles. $\xi=+1$ for boson partition function, while $\xi=-1$  for fermion partition function. In addition, $\{\textbf{r}\}$ denotes $\{\textbf{r}_1,\cdots,\textbf{r}_N\}$. $P$ is a number defined to be the minimum number of times for which pairs of indices must be interchanged to recover the original order $\{\textbf{r}\}$ from $p\{\textbf{r}\}$. With the technique of path integral \cite{barker,Tuckerman,cazorla}, the partition function $Z(\xi,\beta)$ can be mapped as a classical system of interacting ring polymers. In recent works by Hirshberg et al. \cite{HirshbergFermi}, a recursion formula is found to calculate the partition function for both bosons ($\xi=1$) and fermions ($\xi=-1$). 

Assume there are $N_\uparrow$ fermions in state $|\uparrow>$ and $N_\downarrow$ fermions in state $|\downarrow>$, the parametrized partition function for two-component identical particles is
\begin{equation}
Z(\xi,\beta)\sim\sum_{p_1\in S_{N_\uparrow}}\sum_{p_2\in S_{N_\downarrow}}\xi^{P_1}\xi^{P_2}\int d\textbf{x}_1d\textbf{x}_2\cdots d\textbf{x}_N\left<p_1p_2\{\textbf{x}\}|e^{-\Delta\beta\hat H}\cdots e^{-\Delta\beta\hat H}|\{\textbf{x}\}\right>.
\label{twocomponent}
\end{equation}
Here $N=N_\uparrow+N_\downarrow$. ${\textbf x}_j$ denotes both the spatial state and internal state $s_j$ ($|\uparrow>$ or $|\downarrow>$) of the $j$th particle. $p_1$ ($p_2$) is the permutation of fermions with internal state $|\uparrow>$ ($|\downarrow>$). $P_1$ and $P_2$ are the number for permutations $p_1$ and $p_2$, respectively. The recursion formula given in Ref. \cite{Xiong4} for spinor bosons can be generalized to this situation without too much extra difficulty. Path integral molecular dynamics is performed by defining the potential function as
\begin{equation}
U_{\xi}^{(N)}=-\frac{1}{\beta}(\ln W_{\xi}^{(N_\uparrow)}+\ln W_{\xi}^{(N_\downarrow)})+\frac{1}{P}\sum_{j=1}^P V(\textbf{x}_1,\textbf{x}_2,\cdots ,\textbf{x}_N),
\end{equation}
where $V$ is the interaction potential, $W_{\xi}^{(N_\uparrow)}$ may be evaluated recursively as
\begin{equation}
W_{\xi}^{(N_\uparrow)}=\frac{1}{N_\uparrow}\sum_{k=1}^{N_\uparrow}\xi^{k-1}e^{-\beta E_{N_\uparrow}^{(k)}}W_{\xi}^{(N_\uparrow-k)}.
\end{equation}

\begin{equation}
E_{N_\uparrow}^{(k)}=\frac{1}{2}m\omega_{P}^{2}\sum_{l=N_\uparrow-k+1}^{N_\uparrow}\sum_{j=1}^{P}\left(\textbf{x}_{l}^{j+1}-\textbf{x}_{l}^{j}\right)^{2}.
\end{equation}
Here $\textbf{x}_{l}^{P+1}=\textbf{x}_{l+1}^{1}$, except for $l=N_\uparrow$
for which $\textbf{x}_{N_\uparrow}^{P+1}=\textbf{x}_{N_\uparrow-k+1}^{1}$. In addition, $\omega_{P}=\sqrt{P}/\beta\hbar$. Similarly for $W_{\xi}^{(N_\downarrow)}$.

\section{Model}

As a principal demonstration of our algorithm, we consider here two-component Fermi gases in a three-dimensional harmonic trap. We use an attractive Gaussian interaction, because it is convenient for the simulation with path integral molecular dynamics adopted in this work, without the loss of generality.

We consider the following spin-independent Hamiltonian operator
\begin{equation}
\hat H=-\frac{\hbar^2}{2m}\sum_{j=1}^N \Delta_j+\sum_{j=1}^N\frac{1}{2}m\omega^2 |\textbf{r}_j|^2+V_{int},
\end{equation}
with the Gaussian interaction potential
\begin{equation}
V_{int}=-\frac{V}{2w^2}\sum_{j=1}^{N_\uparrow}\sum_{{j'}=1}^{N_\downarrow}e^{-\frac{|\textbf{r}_j-\textbf{r}_{j'}|^2}{w^2}}.
\end{equation}

We consider the length unit $l_u=\sqrt{\hbar/m\omega}$ and energy unit $E_u=\hbar\omega$, and assume $
w=\gamma l_u$.
In this case, the dimensionless Hamiltonian operator is
\begin{equation}
\hat{\tilde H}=-\frac{1}{2}\sum_{j=1}^N\tilde\Delta_j+\frac{1}{2}\sum_{j=1}^N\tilde{\textbf r}_j^2-\sum_{j=1}^{N_\uparrow}\sum_{j'=1}^{N_\downarrow} \frac{\tilde V}{\gamma^2} e^{-|\tilde{\textbf r}_j-\tilde{\textbf r}_{j'}|^2/\gamma^2}.
\end{equation}
Here $\tilde V=V/(\hbar^2/\mu)$ with $\mu=m/2$ the reduced mass for two particle system. 

In free space, if only two fermions with different internal state are considered, the Hamiltonian operator for the relative motion is
\begin{equation}
\hat H_r=-\frac{\hbar^2}{2\mu}\Delta-\frac{V}{2w^2}
e^{-{r}^2/{w^2}}.
\end{equation}
From this Hamiltonian, it is shown in Ref. \cite{Peter} that the s-wave scattering length (in unit of $l_u$) between two particles can be approximated well as
\begin{equation}
\tilde a_s\approx \sum_{i=1}^n\alpha_i\frac{\gamma\tilde V}{\tilde V-W_i}.
\end{equation}
In this work, we consider $n=2$ and use the numerical results $\alpha_1=1.120$, $\alpha_2=0.378$, $W_1=2.684$ and $W_2=17.796$. Because we consider interaction between fermions with different internal states, there exists effective s-wave collision.

\section{Results}

\begin{figure}[htbp]
\begin{center}
 \includegraphics[width=0.75\textwidth]{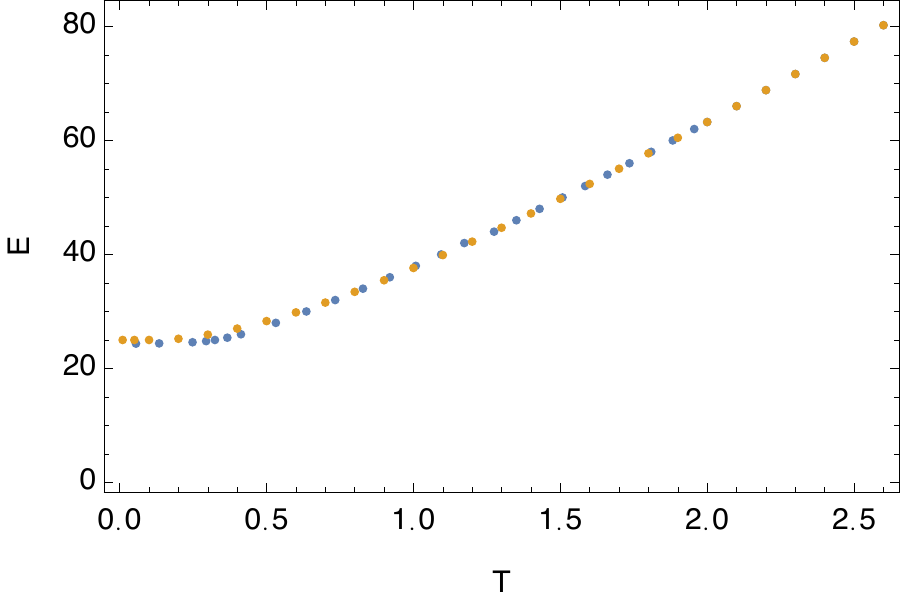} 
\caption{\label{ideal}  For $N_\uparrow=N_\downarrow=5$ Fermi system in a three-dimensional harmonic trap without interaction, the orange circles show the energy of fermions based on the analytical equations (\ref{chemical}) and (\ref{energyc}) in grand canonical ensemble, while the blue circles are obtained with our method, by the determination of the $\xi_E(T)$ curve from the calculated energy of $\xi\geq 0$.}
\end{center}
\end{figure}

To test our algorithm, we first consider $N_\uparrow=N_\downarrow=5$ noninteracting Fermi system in three-dimensional harmonic trap. In Fig. \ref{ideal}, we give the results (orange circles) based on analytical equations (\ref{chemical}) and (\ref{energyc}), and the results (blue circles) of our method based on the approximate $\xi_E(T)$ expression given by Eq. (\ref{xiexample}), determined by the energy data of $\xi\geq 0$. We do find good agreement with each other.

Now we turn to consider the Fermi system with attractive interaction. As an example of our method, we use $N_\uparrow=N_\downarrow=5$, $\gamma=1$ and $\tilde V=2.684$ to consider the thermodynamics of two-component finite-range attractive Fermi gases. The choice of these parameters has the special interest that the s-wave scattering length is divergent. The divergent s-wave scattering length means that what we study is a strongly correlated quantum system, where the exact numerical simulation is needed in principle.

\begin{figure}[htbp]
\begin{center}
 \includegraphics[width=0.75\textwidth]{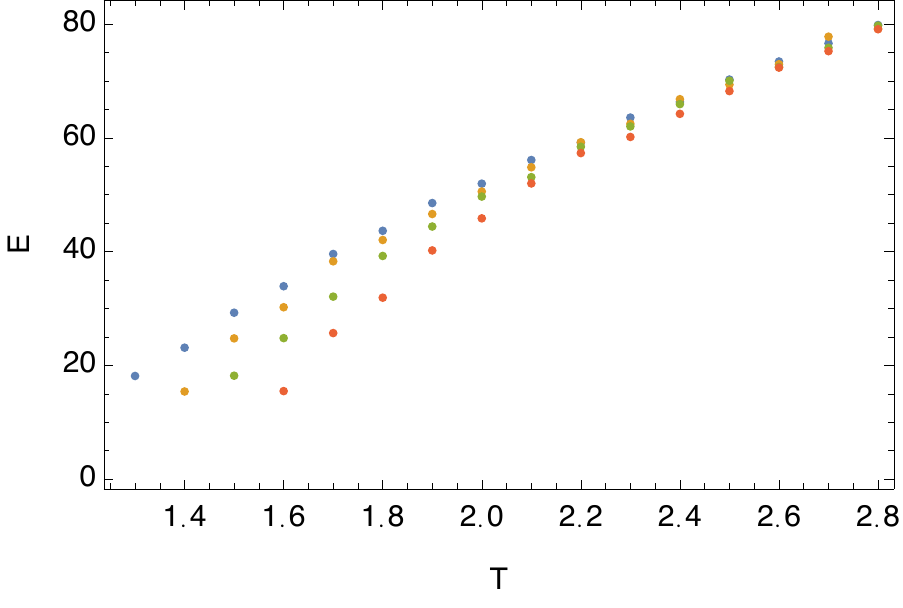} 
\caption{\label{allenergy}  For $N_\uparrow=N_\downarrow=5$ in a three-dimensional harmonic trap with finite-range attractive interaction, shown are the energy for $\xi=0$ (blue circles), $\xi=0.25$ (orange circles), $\xi=0.5$ (green circles), and $\xi=1$ (red circles), respectively.}
\end{center}
\end{figure}

In the presence of attractive interaction, in Fig. \ref{allenergy}, we give the simulation result of the energy for $\xi=0$, $\xi=0.25$, $\xi=0.5$, and $\xi=1$ without suffering from fermion sign problem, respectively. The monotonic behavior of $E(\xi, T)$ for both $\xi$ and $T$ is shown clearly in the simulation result. For details to calculate the energy for $\xi\geq 0$, one may refer to Ref. \cite{Xiong-xi} about the application of separate Nosé-Hoover thermostat \cite{Nose1,Nose2,Hoover,Martyna,Jang} for molecular dynamics simuation.

From all these data for energy, we use polynomial function for interpolation and fitting, so that we get four functions $f_1(T)$, $f_2(T)$, $f_3(T)$ and $f_4(T)$ for the situations of $\xi=0$ (blue circles), $\xi=0.25$ (orange circles), $\xi=0.5$ (green circles), and $\xi=1$ (red circles), respectively.

\begin{figure}[htbp]
\begin{center}
 \includegraphics[width=0.75\textwidth]{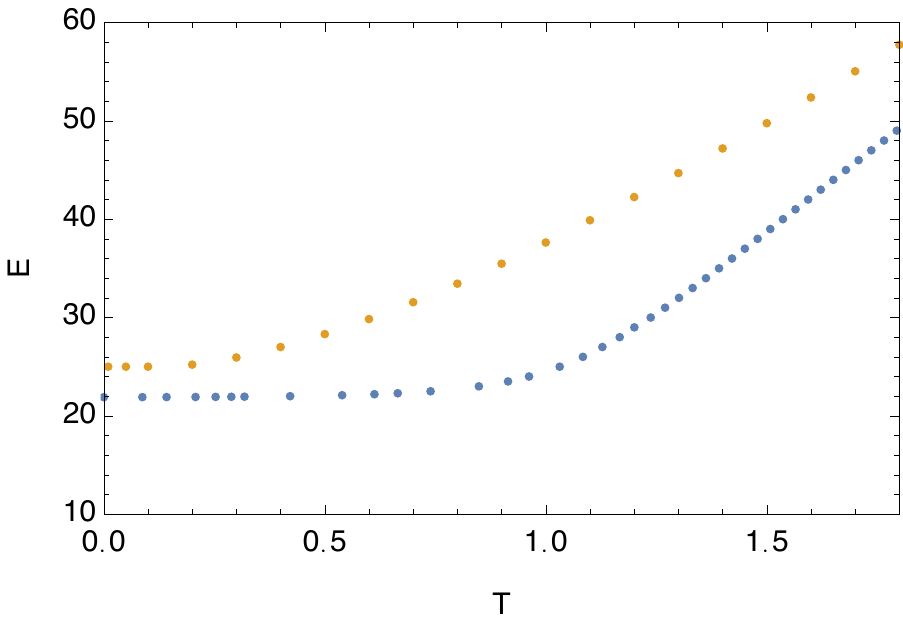} 
\caption{\label{Fermienergy}  For $N_\uparrow=N_\downarrow=5$ in a three-dimensional harmonic trap, the blue circles give the energy of fermions with attractive interaction, while the orange circles show the fermions without interaction.}
\end{center}
\end{figure}

In this example, we use the expression of the $\xi_E(T)$ orbit for constant energy given by Eq. (\ref{xiexample}).
For an energy $E$  larger than the ground state energy of fermions, from the solution of $E=f_j(T)$ ($j=1,2,3,4$), we get four different temperatures $T_1,T_2,T_3,T_4$. In this case, the coefficients $a(E),b(E),c(E),d(E)$ can be obtained from the following four equations:
\begin{equation}
0=b(E)+c(E)T_1^2+d(E)T_1^3,
\nonumber
\end{equation}
\begin{equation}
0.25+0.25^2 a(E)=b(E)+c(E)T_2^2+d(E)T_2^3,
\nonumber
\end{equation}
\begin{equation}
0.5+0.5^2 a(E)=b(E)+c(E)T_3^2+d(E)T_3^3,
\nonumber
\end{equation}
\begin{equation}
1+1^2 a(E)=b(E)+c(E)T_4^2+d(E)T_4^3.
\nonumber
\end{equation}
After we solve $a(E),b(E),c(E),d(E)$ from the above four equations, by setting $\xi_E=-1$ in Eq. (\ref{xiexample}) with these coefficients, we get the temperature $T$ for the fermions with a given energy $E$ in advance. Repeating these simple calculations, we get the energy of fermions with different temperatures, shown in Fig. \ref{Fermienergy}.

\begin{figure}[htbp]
\begin{center}
 \includegraphics[width=0.75\textwidth]{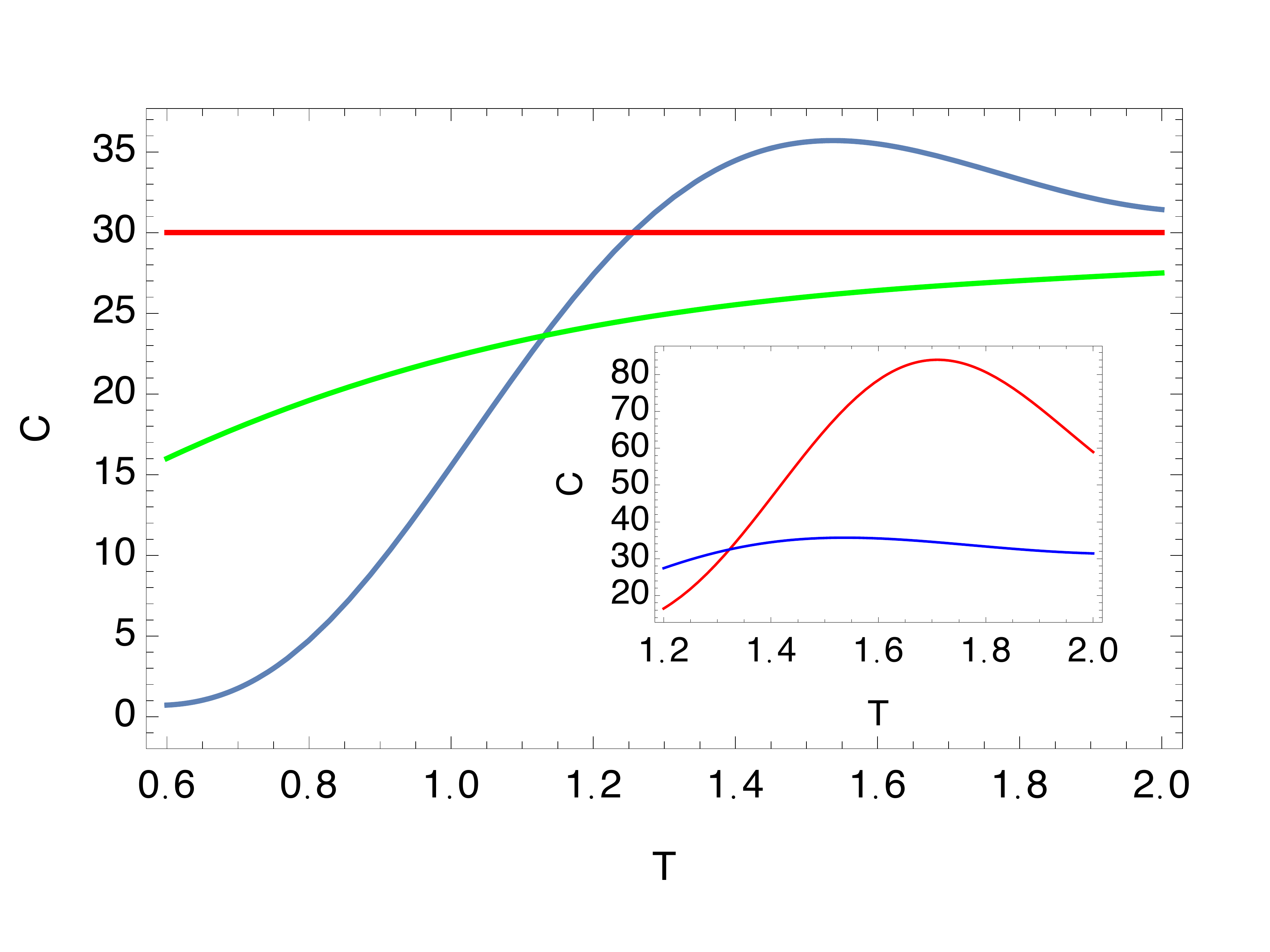} 
\caption{\label{heat}  For Fermi system with $N_\uparrow=N_\downarrow=5$ in a three-dimensional harmonic trap, the blue line gives the heat capacity of the fermions with attractive interaction, while the green line shows the heat capacity for fermions without interaction in grand canonical ensemble. As a comparison, the red horizontal line gives the heat capacity in the classical limit. In the inset of this figure, the blue line shows the heat capacity of the Fermi system with attractive interaction, while the red line shows that of the corresponding Bose system.}
\end{center}
\end{figure}

From the calculated energy, in Fig. \ref{heat}, we give the heat capacity $C(T)=dE(T)/dT$ for fermions with attractive interaction by blue line. We see clearly a peak in the heat capacity at temperature $T=1.54$ for attractive situation. As a comparison, the green line gives the heat capacity of the noninteracting Fermi system in grand canonical ensemble, which has no peak at all. At high temperature, both situations approach the heat capacity in the classical limit, shown by the horizontal red line. The Fermi temperature of the noninteracting system is $T_F=3.48$. 

In the inset of Fig. \ref{heat}, the red line shows the heat capacity of two-component bosons with the same attractive interaction, while the blue line is the result of Fermi system. We see that there is a clear peak in the situation of bosons, too. The temperature of the peak is $T=1.71$ for Bose system, which is larger than that of the corresponding Fermi system. It is clear that the peak in the attractive two-component Fermi system is due to the pairing of the fermions with different spin and the Bose condensation of fermionic pairs. We emphasize that this peak as a demonstration of the phase crossover is due to the many body effect, because for the parameters we choose, the binding energy for $N_\uparrow=N_\downarrow=1$ is zero. This is also verified by our numerical simulation with $N_\uparrow=N_\downarrow=1$ by path integral molecular dynamics. We also notice that the peak of the Bose system is much higher and narrower than the corresponding Fermi system. This is due to the fact that with our parameters, the fermionic pairs are loosely coupled, so that the Pauli exclusion for the fermions in the same spin still plays an important role.

It is worth pointing out that, to map all the energies of fermions, we only use the energy data of $\xi\geq 0$ with $T>1.3$, and in particular $T>1.6$ for the energy data of bosons. This is due to the duality property between Fermi system and Bose system, emphasized in this work.

As a comparison, for $T=1$, we calculate the energy with the previous method in Refs. \cite{XiongFSP,XiongPara}. The idea of analytical continuation along constant temperature predicts $E(\beta,\xi=-1)\approx 29.4$ for two-component Fermi gases with attractive interaction. Because of the potential inflection point below $\xi=0$, this method predicts the energy larger than the actual energy. The present method gives $E=24.6$ which verifies this conjecture, which also supports the effectiveness of our new method in this work. It is hard for us to predict the energy of fermions for $T<1$ along constant temperature, while the present work gives reliable and efficient calculations of the energy for fermions from zero temperature to high temperature.

The present work finally gives the positive answer to the long standing question that whether the transition to Cooper pairs and condensation of Cooper pairs can be simulated from the first principle. The difficulty to answer this question is of course due to fermion sign problem, which is even more severe than the situation of repulsive interaction. In the background of the simulation of the BCS-BEC crossover, a similar peak is found for the case of contact interaction by path integral Monte Carlo, where fermion sign problem is avoided \cite{Burov1,Burov2}.

In all our results in this work, the error due to the statistical fluctuations is negligible for the energy data of $\xi\geq 0$, hence we do not give the error bar in the present work. It is worth pointing out that the main purpose of the present work is to propose and verify the idea of our algorithm. Hence, we provide only the example of $N_\uparrow=N_\downarrow=5$ Fermi system, so that one may follow our calculation more easily. Different from the repulsive interaction studied in Ref. \cite{Xiong-xi}, when more particles are considered, one should be careful that the ground state energy of fermions may correspond to the system with $\xi\geq 0$ in the critical region. In this case, much more molecular dynamics (MD) steps in numerical simulation should be used to try to avoid the difficulty due to the critical slowing down. In addition, the energy data of $\xi\geq 0$ with different temperature should be more dense in this case.
In practical applications, to satisfy the high precision calculation of some problems, we may always consider to increase significantly the MD steps and the number of beads $P$ per particle to satisfy our request. Fortunately, the large-scale computing has maken $N=128$ and $P=256$ feasible by path integral molecular dynamics for bosons \cite{Deuterium}.

\section{conclusion}

As a summary, in this work we extended our previous method on numerical simulation of the thermodynamics of spin polarized fermions, to consider the thermodynamics of two-component fermions with attractive interaction. We ran the simulation for a system of two-component fermions with divergent s-wave scattering length and found the complete heat capacity curve, where a peak is observed signifying the phase crossover caused by fermionic pairing effect. This work verifies the applicability of our new method for the case with attractive interaction, it can be expected the method presented here will find more applications in quantum many body systems, in particular the field of BCS-BEC crossover \cite{Bloch,Randeria}. Even for a few two-component fermions demonstrated in this paper, it may have application to ultracold Fermi atomic gases in optical lattices \cite{lattice}, where each lattice can have a few fermions with different spin. The present work suggests that the measurement of the heat capacity may have the chance to show the fermionic pairing effect by sufficiently decreasing the temperature of the system.

\begin{acknowledgments}
This work is partly supported by the National Natural Science Foundation of China under grant numbers 11175246, and 11334001. 
\end{acknowledgments}

\textbf{DATA AVAILABILITY}

The data that support the findings of this study are available from the corresponding author upon reasonable request. The code of this study is openly available in GitHub (https://github.com/xiongyunuo/PIMD-Pro-2).

\end{document}